\documentclass{comsoc2018}
\usepackage{amsmath,amsthm,amssymb}
\usepackage{mathrsfs,amsmath,wasysym}
\usepackage[ruled,vlined]{algorithm2e}

\usepackage{tikz}
\usetikzlibrary{arrows,%
                petri,%
                topaths,calc,decorations.pathreplacing}
\usepackage{tkz-berge}
\usepackage[position=top]{subfig}
\tikzset{
  font={\fontsize{14pt}{12}\selectfont}}

\renewcommand{\vec}[1]{\boldsymbol{#1}}

\DeclareMathOperator*{\argmax}{arg\,max}
\def\truth{\texttt{Truth}}
\def\KP{\texttt{Prag}}
\def\BR{\texttt{BR}}
\def\CV{\texttt{CV}}
\def\LD{\texttt{LD}}

\def\NN{\texttt{NN}}
\def\BL{\texttt{TMG}}
\def\AVU{\texttt{AU}}

\newtheorem{theorem}{Theorem}
\newtheorem{example}[theorem]{Example}

\usepackage{color}
\usepackage{graphicx}

\newcount\Comments  
\definecolor{darkgreen}{rgb}{0,0.6,0}
\newcommand{\kibitz}[2]{\ifnum\Comments=1{\textcolor{#1}{#2}}\fi}
\newcommand{\rmr}[1]{\kibitz{blue}{[RM:#1]}}
\newcommand{\kg}[1]{\kibitz{red}{[KG:#1]}}
\newcommand{\rf}[1]{\kibitz{green}{[RF:#1]}}
\newcommand{\al}[1]{\kibitz{purple}{[AL:#1]}}

\title{Predicting  Strategic Voting Behavior with Poll Information}
\author{Roy Fairstein, Adam Lauz, Kobi Gal, Reshef Meir}

\begin{document}
\begin{abstract}
The question of how people vote strategically under uncertainty has attracted much attention in several disciplines. Theoretical decision models have been proposed which vary in their assumptions on the sophistication of the voters and on the information made available to them about others' preferences and their voting behavior. 
This work  focuses on modeling strategic voting behavior under poll information. It  proposes  a new heuristic for voting behavior  that weighs the   success of each candidate according to the poll score   with  the utility of the candidate given the voters' preferences. The model weights can be tuned individually for each voter. We compared this model with other relevant voting  models from the literature on data obtained from a recently released large scale study. We  show that the new heuristic outperforms all other tested models. The prediction errors of the model can be partly explained due  to inconsistent voters that  vote for (weakly) dominated candidates. 
 


\end{abstract}

\section{Introduction}
It is well accepted that people often vote strategically in political and other situations, taking into account not just their preferences, but also beliefs about how  their vote would affect the outcome~\cite{Forsythe1996ThreeCandidateExperiments,Bassi2008StrategicManipulationExperimentalStudy}. 
Researchers in economics, political science, and more recently in the computational social choice community, suggested various mathematical models to capture the strategic decision that a voter faces~\cite{conitzer2011dominating,Chopra:04,grandi2013restricted}.
The gains from a certain action depend not only on the preferences of the voter, but also on the votes of others. Thus, part of the difficulty in predicting a voter's decision arises from the fact that there is  \emph{uncertainty} about others' voting decisions, i.e., what can be inferred from the poll about the actual votes.  Theoretical models describe   this uncertainty in the following different  ways, which can lead to   different predictions of a voter's actions.  
\begin{itemize}
\item Expected utility maximization. A rational voter maximizes her expected utility with respect to a \emph{probability distribution} over the actions of the other voters. The distribution itself may be given exogenously (e.g., by a poll with known variance as in our model), or derived via equilibrium analysis from the uncertain preferences of the other voters. Such models were developed mainly in the economics literature and are sometimes known as the ``calculus of voting''~\cite{riker1968theory,merrill1981strategic,MW93}.
\item Voting heuristics. In these models  the voter uses some (typically simple) function that states which action to use at any given situation. The voter is not assumed to be rational, and may not even have a cardinal utility measure or an explicit probabilistic representation of the different outcomes.  For example, according to the 2-pragmatist  heuristics,   the voter behaves as if only the two candidates leading the poll are participating~\cite{RE12}. 
\item Bounded rationality. These models present a mid-point between utility maximization and heuristics. The voter makes a rational strategic decision based on a  heuristic belief, rather than accurate probabilistic belief. One example of such a model is \emph{local dominance}~\cite{MLR14}, which assumes that each voter derives a set of possible outcomes based on a poll, and then selects a non-dominated action within these outcomes. 
\end{itemize}

We evaluated the different models on data obtained from Tal et al.~\cite{TalMG15} who implemented  several voting scenarios in controlled experiments involving humans in different scenarios that vary the number of voters, the poll information, and voters' preferences. 
Our main findings  are that  the $\AVU$ model outperforms all other models in all scenarios.  The $k$-pragmatist heuristic model which considers only a limited number of candidates when making decisions comes in second.  The bounded-rational models obtained  the worst performance.

\subsection{Contribution}
Our first contribution is an extensive evaluation of various decision models  on real-world data.  We use the data of Tal et al.~\cite{TalMG15}, where human subjects with dictated preferences are exposed to a poll over three candidates and make a single voting decision under the Plurality rule. This is the simplest possible setting that involves a nontrivial strategic decision. 

This is the first time that these models are tested versus voting decision with poll information. In fact, for some of them this is the first empirical test at all. 

 Our second contribution is  new heuristic voting rule, inspired by a similar model of Bowman et al.~\cite{bowman2014potential}, that takes into account both the utility of a candidate and its attainability. The \emph{Attainability-Utility} ($\AVU$) decision model  outperforms all other decision models we tested in predicting human votes. 
 
 This contributes to the understanding of the factors that determine  people's strategic voting   and can lead to new theories of voting behavior that combine rational and boundedly rational behavior. 
 



\subsection{Related Work} 
We are not aware of another controlled experiment where voters face multiple strategic decisions with poll information. Yet, similar experiments were conducted in which groups of human players voted strategically with dictated preference profiles. 

Closest to our work is a recent paper by Tyszler and Schram~\cite{tyszler2016information}, who  showed that  the strategic behavior of voters in the lab using  is consistent with a quantal best-response equilibrium.
  The main difference is that their subjects played a strategic game versus other human players, and the information they had was the \emph{preferences} of other voters rather than   poll information. 
 Similar game-theoretic experiments along that line were conducted in \cite{Forsythe1996ThreeCandidateExperiments,Bassi2008StrategicManipulationExperimentalStudy,van2010strategic}. In particular, these studies have shown that strategic voting in the lab increases with the amount of information subjects received about  others' preferences and  actions.
 
 
 A different line of works in political science compared theoretical models against   actual votes in political elections (using exit polls to obtain the truthful preferences). For example, Blais et al.~\cite{blais2000calculus} tested the calculus of voting model on empirical data from political elections and focused on voter's decision to vote/abstain. They concluded that the model has some explanatory power but is far from explaining the data completely, and did not compare to other decision models. In contrast, Abramson et al.~\cite{abramson1992sophisticated} concluded that voting behavior in US primary elections is consistent with the calculus of voting, but observed obvious strategic behavior only in $\sim$13\% of the voters, a bit higher than the fraction that seem to vote at random.
 
In contrast to controlled experiments, such datasets typically contain few decisions of each voter (usually just one), and are this insufficient to test decision models versus individual behavior.  

\section{Preliminaries}

\def\calU{\mathcal {U}}
\def\calD{\mathcal {D}}
In this section we provide the necessary background for our work. 
An (anonymous)  \emph{score aggregation rule} with $m$ candidates $C$ is a function $f:\mathbb N^m\rightarrow 2^C\setminus\emptyset$, mapping vectors of candidates' scores to a subset of winning candidates. In particular, the Plurality rule lets each voter vote for a single candidate, collects the total number of votes $s\in \mathbb N^m$, and selects $f(s)=\argmax_{c\in C}s(c)$. 

We consider a single voter who faces a decision, to vote for one of several candidates $C$. The voter has a cardinal utility function $u:C\rightarrow \mathbb R$, where $u_i(c)$ is the utility of the voter if candidate $c$ wins (different utility for each candidate). The utility of a subset of winners $W\subseteq C$ is $u(W)=\frac{1}{|W|}\sum_{c\in W}u(c)$. 
Denote by $\calU(C)$ the set of all utility functions over the set $C$. We denote by $f(s+c)$ the outcome of the score vector $s$ with one additional vote to $c$. 

Prior to her vote, the voter is faced with poll information, which is a point estimate of candidates scores under the Plurality voting rule. Formally, the poll is a vector $\vec s\in \mathbb N^m$, where $s(c)$ is the number of voters expected to vote for candidate $c$. There is a joint probability distribution $\calD\in \Delta(\mathbb N^m \times \mathbb N^m)$ over pairs of ``real outcomes'' and polls.\footnote{A priori, this distribution could be arbitrary, but in most realistic cases there is some correlation between the real score of a candidate and its score in the poll.} The voter is not explicitly informed of this distribution.

\def\eps{\varepsilon}
\medskip
A \emph{decision model} (for Plurality with $m$ candidates and a poll) is a function $M:  \calU(C) \times \mathbb N^m \rightarrow C$, where $M(u,\vec s)\in C$ is the vote of a voter with utility function $u$, using decision model $M$ given a poll $\vec s$.  We use a superscript for the name of the decision model (e.g., $M^\truth$), and subscripts to denote voter-specific parameters, if relevant. We restrict our attention to deterministic decision models in this work.
 We demonstrate with two simple examples.
First, the decision model of a voter who is always truthful regardless of the poll is $M^\truth(u, \vec s) = \argmax_{c\in C}u(c)$. 

Next, consider a rational voter that believes the poll to be a completely accurate representation of the other votes. Such a voter can predict that the outcome of voting $c$ is $f(\vec s+c)$, and her decision will be   $M^{\BR}(u,\vec s)\in \argmax_{c\in C}u(f(\vec s+c))$, i.e., a ``best response'' to the votes of the other voters (with some assumption on how to vote when there are multiple best responses). 

\begin{figure}[t]
\centering
\subfloat{{\includegraphics[width=7cm]{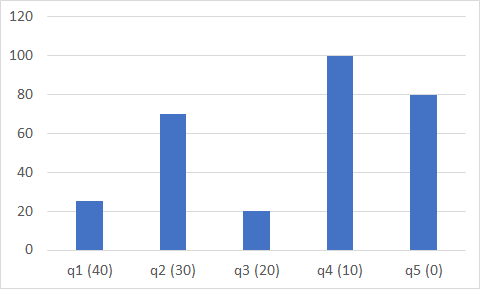} }}
\qquad
\subfloat{\begin{tabular}{l|c}
   Decision model   & vote\\
      \hline
      $\KP, \quad k=2$& $q_4$\\
      $\KP, \quad k=4$& $q_1$\\
      $\CV, \quad \eta=8$& $q_2$\\
      $\CV, \quad \eta=10000$& $q_4$\\
          $\LD,\quad r=0.01$& $q_1$\\
      $\LD, \quad r=0.08$& $q_2$\\
      \hline
         $\LD$ + LB, $r=0.01$& $q_4$\\
      $\LD$ + LB, $r=0.08$& $q_2$\\
    \end{tabular}}
    \caption{\label{fig:pollExample}The poll $\vec s$ from Example~\ref{ex:poll}, and the candidate selected by each decision model.}
\end{figure}

For exposition, we introduce a running example with 5 candidates, and specify which candidate the voter will choose under every decision model. 
\begin{example}[Running example]
\label{ex:poll}
The set of candidates is $C=\{q_1,\ldots,q_5\}$. A voter's utility is described by the vector $u=(40,30,20,10,0)$ (preferences are lexicographic). Poll scores are given by $\vec s=(25,70,20,100,80)$.
\end{example} 
Figure~\ref{fig:pollExample} shows the scores of all candidates graphically. 
  Both $M^\truth$ and $M^\BR$ always select $q_1$.

\subsection{Decision models from the literature}\label{sec:lit_models}
 In this section we briefly describe some decision-making models of voting behavior  from the literature, one for each of the approaches specified above (heuristic, rational, and bounded-rational). In Section~\ref{sec:auv} we describe our decision model developed for this study, and in Section~\ref{sec:methodology} we provide a detailed comparison of this model to the decision-making models below as well as several baseline models.

\paragraph{k-pragmatist}The first model we consider is the simple $k$-pragmatist heuristic~\cite{RE12}. Formally, let $B_k(\vec s)$ contain the $k$ candidates with highest score in $\vec s$, then the pragmatist decision model with parameter $k$ selects the most preferred candidate among them, i.e.,
$$M^\KP_k(u,\vec s) = \argmax_{c\in B_k(\vec s)}u(c).$$
We allow $k$ to be an individual parameter that differs from voter to voter.

In Figure~\ref{fig:pollExample} we see that for $k=2$ the voter look only at the two leading candidates ($q_4$ and $q_5$), and therefore will vote to the one that is more preferred among them. For $k=4$, the voter considers all  candidates except $q_3$ as possible winners, and therefore will vote for his most preferred candidate $q_1$.

\paragraph{Calculus of voting}
The calculus of voting suggests that a voter always votes in a way that maximizes her expected utility~\cite{riker1968theory,MW93}. The complications of the model usually arise from the fact that the voter is assumed to know only the other voters' \emph{preferences}, and uses an equilibrium model to predict their votes. However,  we consider a simpler version where the distribution of votes is given exogenously~\cite{merrill1981strategic}.

Recall that we defined $\calD$ as a joint distribution over actual scores and polls. We denote by  $\calD(\vec s)$ the  distribution on the actual scores, conditional on poll scores  $\vec s$. 
Denote by 
$$
P_{\vec s,\calD}(x,y)=Pr_{\vec s'\sim \calD(\vec s)}\left[
\begin{array}{l}
(f(\vec s')=\{x\} \text{ and } f(\vec s'+y)=\{x,y\}) \text{   or   }\\
(f(\vec s')=\{x,y\} \text{ and } f(\vec s'+y)=\{y\})
\end{array}
\right],
$$
the probability that the voter is \emph{pivotal} for $y$ versus $x$ when the poll is $\vec s$. That is, voting for $y$ makes $y$ a joint or unique winner. Then, the voter votes so as to maximize her expected utility:
$$M^{\CV}_\calD(u,\vec s) = \argmax_{c\in C}\sum_{c'\neq c}P_{\vec s,\calD}(c',c)(u(c)-u(c')).$$ 

\medskip
To make the CV model concrete, we need to pin it down to a specific distribution $\calD$. For this paper, we use the way that scores were generated  in the experiment of Tal et al.~\cite{TalMG15}. Specifically, 
given poll $\vec s$ of a $n$-voter population, the actual score vector $\vec s'$ is obtained by sampling $n$ votes from a multinomial distribution whose parameters are $(\frac{s(c)}{n})_{c\in C}$. 

We use $P_{\vec s,\eta}$ as a shorthand for $P_{\vec s,\calD}$ when $\calD(\vec s)$ is a multinomial distribution with $\eta$ voters  as explained above.
 When $\eta=n$ (i.e., the true number of voters), this means that $M^{\CV}_\eta$ selects the candidate that exactly maximizes the voter's expected utility, since  $P_{\vec s,n}(x,y)$ is the true probability that the voter is pivotal.

However, the $M^\CV_\eta$ decision model allows for a more flexible, bounded-rational decision: when $\eta<n$ the voter overestimates her  true pivot probability, and thus her influence on the outcome, whereas $\eta>n$ means that the voter underestimates her influence.

In Figure~\ref{fig:pollExample} we see that for $\eta=8$ the voter  believes she is pivotal with sufficiently  high probability to substantially increase the chance of $q_2$ to win.  However for $\eta=10007$, the voter believes that any tie except a tie of $(q_4,q_5)$ is highly improbable, and therefore will vote for $q_4$.

\paragraph{Local dominance}
Under the Local dominance model~\cite{MLR14,Meir15}, the voter has an `uncertainty parameter' $r$. Given poll $\vec s$ with $n$ participants, the voter considers as possible (without assigning any explicit probability) all score vectors $\vec s'$ such that $|\vec s(c)-\vec s'(c)|\leq r\cdot n$ for all $c\in C$. Then, the voter selects an undominated action (i.e., candidate) given this set of possible outcomes. Meir et al.~\cite{MLR14} characterize the undominated candidates:
\begin{itemize}
\item Let $W$ be all candidates whose score in $\vec s$ is at least $\max_{c\in C}s(c)-2rn$.
\item If $|W|\geq 2$, then the undominated candidates are all candidates in $W$ except the least preferred.
\item If $|W|=1$, then all candidates are undominated. 
\end{itemize}
Denote by $U(\vec s,u,r)$ the set of undominated candidates in poll $\vec s$ for a voter with utility $u$ and parameter $r$. The decision model of such a voter is 
$$M^{\LD}_r(u,\vec s) = \argmax_{c\in U(\vec s,u,r)}u(c).$$
This assumes that the voter selects the most preferred undominated candidate, if more than one exists. 

In Figure~\ref{fig:pollExample} we see that for $r=0.01$ the voter believes that the poll is very accurate (the score of each candidate may change by at most $rn=3$ votes), and there is only one possible winner ($W=\{q_4\}$). In this case, the voter remains truthful and $M^\LD_{0.01}(u,\vec s) = q_1$. 
When $r=0.08$, the voter believes that the poll is not very accurate and  hence $W=\{q_2,q_4,q_5\}$. 





\section{The Attainability-Utility Heuristic}
\label{sec:auv}
We suggest a new heuristic $M^\AVU$  that separately evaluates the \emph{attainability} (an approximation of the success of each candidate according to the poll score)   and the utility of the candidate given the voter's preferences. It  selects the candidate that maximizes their weighted geometric mean. The heuristic is partly inspired by a rule that was used in the simulations of Bowman et al.~\cite{bowman2014potential} for multi-issue voting.\footnote{In \cite{bowman2014potential}, attainability was computed for each issue separately and there were additional factors such as learning from the past. All factors were multiplied to obtain the heuristic attractiveness of the candidate.}  


Given a poll $\vec s$, we compute the attainability of each voter $c$ similarly to \cite{bowman2014potential}: 
$$a_\beta(c,\vec s) = \frac{1}{1 + \exp({-\beta \cdot (\frac{s(c)}{n} - \frac{1}{m}))}}\in [0,1].$$

Then, for  some small constant $\eps>0$, we define:
 $$M^\AVU_{\alpha,\beta}(u,\vec s) = \argmax_{c\in C}H^\AVU_{\alpha,\beta}(u,\vec s,c),\text{ where } H^\AVU_{\alpha,\beta}(u,\vec s,c)=\left((\eps+u(c))^\alpha \cdot (\eps+a_\beta(c,\vec s))^{2-\alpha}\right).$$

Intuitively, the $\alpha$ parameter trades-off the relative importance of attainability and utility, where $\alpha=0$ means the voter always selects the candidate with maximal score,\footnote{This is the only case where $\eps>0$ is needed, as $u(c)$ may be $0$.} and $\alpha=2$ means the voter is always truthful.

The $\beta$ parameter can be thought of as the accuracy of the poll in the eyes of the voter,  similarly to the  role of parameter $r$ in the LD model and $\eta$ in the CV model.

  
  


\begin{figure}[t]
\begin{center}
\includegraphics[scale=0.5]{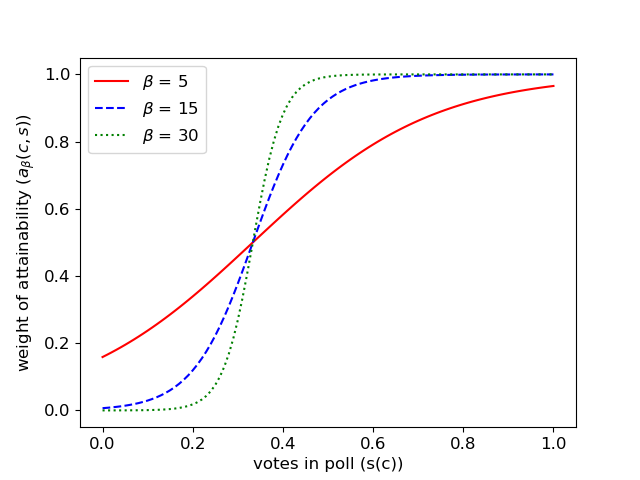}
\caption{\label{fig:pollScore}  Attainability as a function of candidate poll scores for different values of $\beta$.\rmr{use solid/dashed/dotted lines. larger fonts.}\rf{done}}
\end{center}
\end{figure}

  Figure~\ref{fig:pollScore} shows how $\beta$ affects  the attainability score  $a_\beta(s)$. Candidates that are tied have the same attainability. High $\beta$ means that a small advantage in score translates to a large gap in attainability. 

Table~\ref{tab:AvUDecision} 
shows the $\AVU$ model behavior over Example~\ref{ex:poll} with different parameters.
For $\alpha$  close to 2, the voter is tend to be more truthful, however when there is a big gap in votes between candidates, $\beta$ which of the top preferences to choose. There is big gap between $q_1$ and $q_2$, therefore when $\beta$ is big, the gap will cause a bigger difference in the score and will cause to vote for $q_2$. In contrast When $\beta$ is small, this gap is less taken into account, and the voter will vote for $q_1$ . In contrast, when $\alpha$ is close to 0, the model considers the poll as more important: when $\beta$ is large, smaller changes in the poll will have more effect on the decision and therefore only $q_2,q_4$ and $q_5$ have non-negligible attainability;
for small $\beta$ the difference in poll have less effect and therefore preference 4 was chosen. Notice that no matter what the parameters are, the model will never choose to vote for $q_3$ or $q_5$,  since they are each dominated by another candidate with higher score \emph{and} utility.


\rmr{For future: relate better the sigmoid function to probability of winning}

\begin{table}[ht!]
  \begin{center}
    \begin{tabular}{l|c|c|c|c|c|c}
      & $H(q_1)$ & $H(q_2)$ & $H(q_3)$ & $H(q_4)$ & $H(q_5)$ & $M^\AVU_{\alpha,\beta}(u,\vec s)$\\
      \hline
      $\alpha=1.8,\beta=30$ & 382.9 & \textbf{433.3} & 100.1 & 64.0 & 0.01 & $q_2$\\
      $\alpha=1.8,\beta= 10$ & \textbf{578.7} & 413.2 & 162.6 & 61.4 & 0.01 & $q_1$\\
      $\alpha=0.2,\beta=30$ & $\approx 0$ & 1.18 & $\approx 0$ & \textbf{1.54} & 0.51 & $q_4$\\
      $\alpha = 0.2,\beta = 10$ & 0.16 & 0.77 & 0.11 & \textbf{1.06} & 0.3 & $q_4$\\
    \end{tabular}
    \caption{\label{tab:AvUDecision}$M^\AVU_{\alpha,\beta}$ heuristic score and decision in Example~\ref{ex:poll}, for various parameter values.}
  \end{center}
\end{table}

\section{Methodology}
\label{sec:methodology}
\paragraph{Dataset}We evaluated the different models on data obtained from Tal et al.~\cite{TalMG15} who implemented  several voting scenarios in controlled experiments involving humans. 
Some of this data is publicly 
 available at the link \url{votelib.org}.  The data was obtained from 595 distinct subjects. Each subject played up to  20 rounds of voting  with 3 candidates,  each time with different  preferences and poll information.
 The poll provided a noisy indication of the results of the voting.  The voting  instances  can be divided into six different ``scenarios'' corresponding to  different orders of candidates' scores in the poll  once preferences are held fixed (See two leftmost columns in Table~\ref{tab:f_all}). 
 
 We denote the candidate as $Q$ for the most preferred, $Q'$ for the second and $Q''$ for the least preferred. The reward was $10\cent$ for each round where $Q$ was elected, $5\cent$ for $Q'$, and $0\cent$ for $Q''$. Note that only in scenarios E and F, where $Q$ is ranked last at the poll, the voter may have a monetary incentive to  vote for  $Q'$, and never has an incentive to vote for $Q''$. 

\subsection{Evaluated decision models}
In addition to our decision model $M_{\alpha,\beta}^\AVU$, we evaluate the following single-parameter decision models described in Section~\ref{sec:lit_models}: $M_k^\KP,M_\eta^\CV,M_r^\LD$. To these models, we add several other baselines.

\paragraph{Voter type based model}
 
Tal, Meir and Gal~\cite{TalMG15} identified 3 distinct types of voter behavior, albeit without suggesting an explicit decision model:
\begin{enumerate}
\item Voters who are always truthful (TRT voters, about 10\%-15\% of subjects);
\item Voters that often compromise when $Q$ is ranked last (CMP voters, about 40\% of subject), and otherwise vote truthfully;
\item Voters that often compromise AND select the leader $Q'$ when ranked first (LB voters, about 50\% of subjects).
\end{enumerate}
They also identified a subgroup of subjects  who select unjustified  actions (a candidate $c$ where there is $c'$ that is both more preferred and higher-ranked) more than once. The behavior of these voters (about 5\%-10\% of the dataset) is naturally harder to predict for any decision model. We analyze the results for \emph{all subjects}, but return to the issue of unjustified actions and voters in Section~\ref{sec:where}.

Based on their distinction of types, we consider the simple TMG decision model $M^{\BL}_T$. The parameter $T\in \{TRT,CMP,LB\}$ is the voter type. It is defined as follows:
\begin{itemize}
\item $M^{\BL}_{TRT}(u,\vec s)=M^{\truth}(u,\vec s)=Q$; 
\item $M^{\BL}_{CMP}(u,\vec s) = Q'$ if $Q$ is ranked last in $\vec s$,  and $Q$ otherwise;
\item $M^{\BL}_{LB}(u,\vec s) = Q'$ if $Q'$ is ranked first in $\vec s$, and $M^{\BL}_{LB}(u,\vec s)=M^{\BL}_{CMP
}(u,\vec s)$ otherwise.
\end{itemize}

\paragraph{Local-Dominance with Leader bias}
Note that the findings of \cite{TalMG15}  indicate a strong tendency to bias for the leader of the poll, which is not taken into account in the Local dominance model. 
We thus consider a ``leader-biased'' variation of the local dominance model:
$$M^{\LD+LB}_r(u,\vec s) = M^{\LD}_r(u,\vec s) \text{ if }|W|\geq 2,\text{
 and otherwise $M^{\LD+LB}_r(u,\vec s) =W$}.$$
 
 In Figure~\ref{fig:pollExample} we see that this model acts similar to the LD model, however when there is only one possible winner, this model allows the voter to be leader biased and voter for his fourth preference instead of being truthful.

\paragraph{Black-box neural network predictor}
Another baseline we used was a general black-box classifier. We extracted about 30 relevant features, including the poll scores, the differences in poll scores, voter's utility and the voter type as identified in \cite{TalMG15}. The ``decision model'' $M^\NN$ then feeds all features to the classifier, which predicts an action in $C$.

The classifier consisted of a single-hidden-layer feed-forward neural network classifier.
The input nodes represented features that  summarized  voters' preferences, and the poll information that was provided to them, and information about the voter  types. 
The classifier was implemented using the   \texttt{nnet} package of \texttt{R}.\footnote{\url{https://cran.r-project.org/web/packages/nnet/index.html}}

 \subsection{Evaluation metrics}
\paragraph{Prediction and parameter fitting}
The prediction was performed using leave-one-out method. For each voter we excluded one of his rounds, one by one. Using the rest of the rounds we learned the relevant model  parameters and predicted what the voter will do in the excluded round. 

\paragraph{Confusion matrices}
The predictions for a specific decision model result in a \emph{confusion matrix}: The entry $A(x,y)$ in the matrix specifies how many times the model $M^\AVU$ predicted $x$ and the actual voter action was $y$ (a matrix where all off-diagonal entries are 0 indicates perfect prediction). Both rows and columns are sorted $Q,Q',Q''$.
An example for a confusion matrix from out data:
\begin{center}
confusion matrix: $ A=\left[\begin{array}{ccc} 
						  5409 & 441 & 132\\
                          32 & 2538 & 90\\
                          117 & 188 & 373
                          \end{array}\right]$
\end{center}
For example,  in 441 samples (4.7\% of the data), the studied model in the example predicted $Q'$ but the voter selected $Q$.

\paragraph{Performance measures}
From the confusion matrix we compute standard measures for multi-class prediction problems~\cite{powers2011evaluation}.  These include precision, recall as well as the  f-measure, which 
is the harmonic mean of precision and recall,  for every candidate $c\in C$. 
$$\textrm{prec}(c) = \frac{A(c,c)}{Col_A(c)};~~\textrm{recall}(c) = \frac{A(c,c)}{Row_A(c)};~~F(c)=\frac{2\textrm{prec}(c)\cdot\textrm{recall}(c)}{\textrm{prec}(c)+\textrm{recall}(c)},$$
where $Col_A(c) = \sum_{c'\in C}A(c',c), Row_A(c)=\sum_{c'\in C}A(c,c')$.
Since there are three possible actions, we calculate a single f-measure by weighting each f-measure by the number of times this action was played:
$$F_A = \frac{1}{\|A\|_1}\sum_{c\in C}Row_A(c)F(c).$$

In the example matrix above, $\textrm{prec}(Q') = \frac{2538}{441+2538+188}=0.801$, $\textrm{recall}(Q') = \frac{2538}{32+2538+90}=0.954$, and $F(Q') = \frac{2 \cdot 0.954 \cdot 0.801}{0.954+0.801} = 0.87$. For the entire matrix, we would get   
$F_A = \frac{5982 \cdot 0.937 + 2660 \cdot 0.87 + 678 \cdot 0.586}{9320} = 0.892$.

An f-measure of $1$ means that the decision model perfectly explains the data.




\section{Results and Analysis}
Table~\ref{tab:f_all} shows the f-measure of each decision model. We emphasize that the individual parameters of each voter were learned using leave-one-out to avoid overfitting. The results are separated to the different poll scenarios, as they each reflect a different strategic decision. The f-measure is also presented graphically   in the solid bars shown  in Figure~\ref{fig:f_all}. 
\begin{table}[ht!]
  \begin{center}
    \begin{tabular}{l|c|c||c|c|c|c|c|c|c}
      \multicolumn{2}{c|}{scenario} & frequency  & $\AVU$ & $\LD$ & $\LD$+LB & $\CV$ & $\KP$ & $\BL$ & $\NN$\\
      \hline
      A & $Q > Q' > Q''$ & $(15.0\%)$ & 0.902 & 0.902 & 0.902 & 0.902 & 0.902 & 0.902 & 0.904\\
      B & $Q > Q'' > Q'$ & $(11.1\%)$ & 0.903 & 0.903 & 0.903 & 0.903 & 0.903 & 0.903 & 0.909\\
      C & $Q' > Q > Q''$ & $(14.8\%)$ & 0.734 & 0.389 & 0.691 & 0.389 & 0.610 & 0.626 & 0.697\\
      D & $Q'' > Q > Q'$ & $(16.9\%)$ & 0.722 & 0.657 & 0.678 & 0.657 & 0.695 & 0.657 & 0.709\\
      E & $Q' > Q'' > Q$ & $(22.7\%)$ & 0.736 & 0.486 & 0.680 & 0.642 & 0.708 & 0.655 & 0.704\\
      F & $Q'' > Q' > Q$ & $(19.5\%)$ & 0.571 & 0.414 & 0.461 & 0.470 & 0.559 & 0.407 & 0.551\\
      \hline
      \multicolumn{3}{c||}{total} & 0.759 & 0.591 & 0.706 & 0.676 & 0.729 & 0.708 & 0.739\\
    \end{tabular}
    \caption{\label{tab:f_all}f-measure of all tested models, separated by scenario. \rmr{what do you mean by ``combined''?}\rf{irrelevant, changed caption.}}
  \end{center}
\end{table}

\subsection{Main findings}
\begin{figure}[t]
\begin{center}
\subfloat{\includegraphics[scale=0.7]{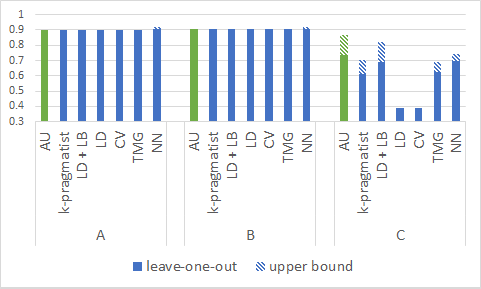}
\includegraphics[scale=0.7]{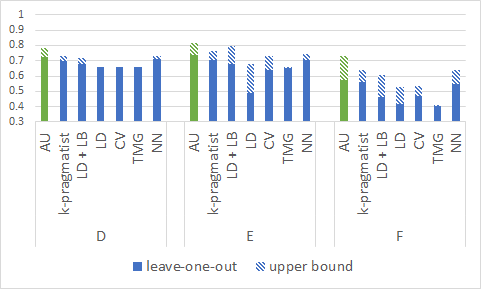}}
\hfill
\includegraphics[scale=0.6]{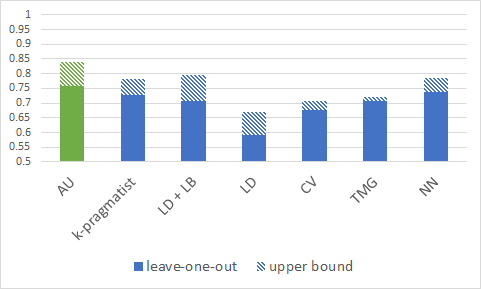}
\caption{\label{fig:f_all} The f-measure of each decision model in all scenarios.  The solid bars show the prediction performance, whereas the stripped bars show the upper bound on the performance of each model. \rmr{make fonts larger}\rf{increased}}
\end{center}
\end{figure}

 From Table~\ref{tab:f_all} and Figure~\ref{fig:f_all} we can derive the following insights:
\begin{itemize}
\item In Scenarios A and B,  all decision models (except $\NN$) predict that voters are always truthful, and thus have the same high performance. 
\item In all scenarios C-F, the $\AVU$ model outperforms all other models.
\item The ``sophisticated'' bounded-rational models $\CV$ and $\LD$ have the worst performance. In particular, they demonstrate poor performance in Scenario~C where voters' decision is influenced by leader-bias~\cite{TalMG15}.
\item The $k$-pragmatist heuristics performs surprisingly well, considering its utter simplicity and the fact that it only allows three types of voters (for $k=1,2$ or $3$). 
\item The LB variant of local dominance strictly improves its performance, placing it roughly at par with $k$-pragmatist and the neural network predictor.
\item Scenario~F is the most difficult one for almost all models, with the best models having an f-measure slightly above $0.5$.
\end{itemize}

\subsection{Upper bound on Performance}
The data we use to fit the parameters of each voter is sparse.  Each voter has at most 20 samples, and in some scenarios only 1 or 2 samples (or none at all). 
Therefore,  even leaving out a single sample may significantly hurt performance. In order to find what is the maximum explanatory power of each model, we re-calculated the f-measure for each model using the entire dataset both as a training set and a test set. 
 Clearly this approach is suspect to overfitting, so it only provides an upper bound on the prediction ability of each model.

These upper bounds appear as stripped bars in Fig.~\ref{fig:f_all}. Note that our $\AVU$ model still outperforms all other models when we compare upper bounds (with the slight exception of the $\NN$ model in Scenarios A and B). The numerical upper bounds can be found in the appendix Table~\ref{tab:upper}.

\subsection{Where are the errors?}\label{sec:where}

Next, we dig deeper into the results. We want to see what kinds of mistakes the $\AVU$ decision model  tends to have. For example,  whether these mistakes concentrate on a specific subset of subjects or scenarios. These insights could be used later to improve the model, and to design further experimental evaluation.

\paragraph{Errors by poll size}First, the model seems to perform equally well for all poll sizes (see Table~\ref{tab:pollSizeFmeasure} in appendix), even as different as $n=8$ and $n=10000$. We thus conclude that the size of the poll is not a major factor in explaining the prediction errors.   

\paragraph{Errors by type}

\begin{figure}[ht!]
  \begin{center}
    \begin{tabular}{cccc}
     
      $C.\quad (Q' > Q > Q'')$  &
	  $D.\quad (Q'' > Q > Q')$ & 
         $E.\quad (Q' > Q'' > Q)$ &
      $F.\quad (Q'' > Q' > Q)$
      \\
      $\left[\begin{array}{ccc}
                          			734 & 153 & 0\\
                          			241 & 465 & 0\\
                          			16 & 7 & 0
                          			\end{array}\right]$ 
                             & $\left[\begin{array}{ccc}
                          			1323 & 0 & 116\\
                          			128 & 0 & 35\\
                         			171 & 0 & 119
                          			\end{array}\right]$ 
                                        &$\left[\begin{array}{ccc}
                          			480 & 348 & 0\\
                          			231 & 1414 & 0\\
                         			13 & 42 & 0
                          			\end{array}\right]$ 
                                   & $\left[\begin{array}{ccc}
                          			487 & 227 & 74\\
                          			275 & 606 & 93\\
                          			78 & 175 & 153
                          			\end{array}\right]$ 
                                   
    \end{tabular}
    \caption{\label{tab:confusionMatrices}Confusion matrix for different scenarios.}
  \end{center}
\end{figure}
Figure~\ref{tab:confusionMatrices} shows the confusion matrices of the $\AVU$ decision model in all the ``interesting'' scenarios C-F. Recall that all the off-diagonal entries indicate prediction errors, where the column is the predicted action ($Q,Q'$ or $Q''$, in that order) and the row is the action of the subject.

As can be seen in the table, most of the prediction errors in Scenario~C are due to under-prediction of voting for the leader $Q'$. In contrast, most of the errors in Scenario~E are due to over prediction of a strategic compromise $Q'$. We can also see why Scenario~F is the hardest, as all three actions are played frequently.

\paragraph{Errors by subject}Every decision model can capture the behavior of some human subjects better than others. To check how well different subjects are predicted, we computed the confusion matrices and f-measure for each of the 595  subjects, when actions are predicted by $M^\AVU$. An f-measure of $1$ means that all actions of this subject were predicted correctly. 

\begin{figure}[h]
\begin{center}
\includegraphics[scale=0.8]{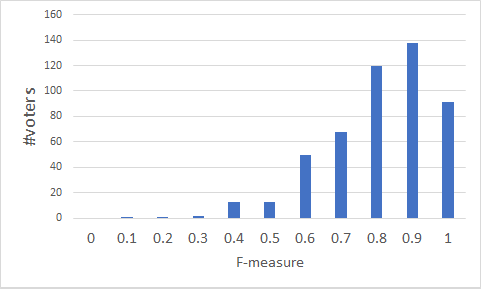}\\
\caption{\label{fig:AvU_f_dist}A histogram showing, for every f-measure $F$, how many subjects have  f-measure of $F$.  }
\end{center}
\end{figure}

Figure~\ref{fig:AvU_f_dist} shows the distribution of subjects' individual f-measures. We can see that about $46\%$ of the subjects are predicted very well (f-measure over 0.9), $29\%$ predicted reasonably well (f-measure over 0.8), and the rest of the subjects (about $25\%$) are with f-measure less than 0.8. 

This means that most of the prediction errors are due to a relatively small subset of  subjects. One possible explanation is that these are the subjects who played fewer games, and therefore it is harder for the model to learn their parameters, however we get a similar distribution after omitting subjects who played under 10 games. 

The main question is thus whether voters whose behavior is not predicted well follow a different decision model than $\AVU$, or are somehow inherently unpredictable.


\paragraph{Inherently inconsistent behavior} 
To answer the above question, we considered types of behavior that would be `inherently unpredictable.'
For example, \cite{TalMG15} categorized as ``unjustified'' a sample $(\vec s,a)$ (action $a\in C$ under poll $\vec s$) if there is another candidate $a'$ that `dominates' the selected action $a$ ($a'$ is more preferred than $a$ and $s(a')\geq s(a)$). They showed that voters with at least two unjustified actions have a random component in their behavior.

We suggest an additional criterion that is  based on inconsistency among a voter's own actions. We say that a sample $(\vec s,a)$ is \emph{inconsistent} if there is another sample $(\vec s^*,a^*)$ of the same voter, such that: (i) $a^*\neq a$; (ii)  $s^*(a)\geq s(a)$; and (iii) $s^*(a') \leq s(a')$ for all $a'\neq a$. In words, $a$ is in a weakly better position in $\vec s^*$, but still the voter prefers to vote for another candidate $a^*$. \kg{why don't you consider preferences of the voter  like the unjustified example of [17]?}\rmr{it is a different kind of inconsistency}

\begin{figure}[t]
\begin{center}
\includegraphics[scale=0.7]{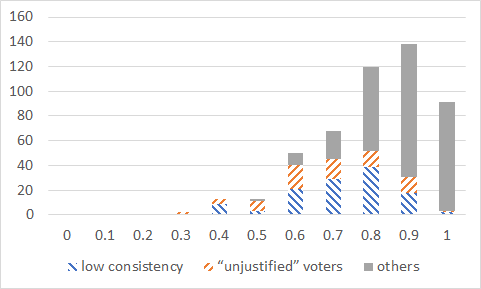}~
\includegraphics[scale=0.7]{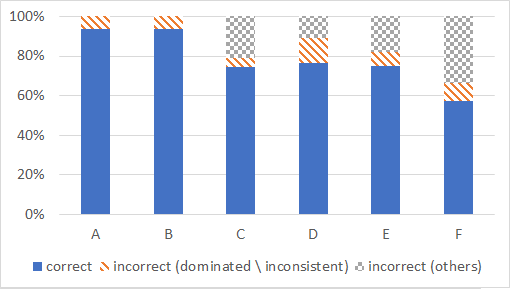}\\
\caption{\label{fig:VotersConsistencyFMeasure}\textbf{Left}: f-measure per voter for voters with more than 10 samples. We divided the population into ``unjustified'' voters who played at least two unjustified actions; ``inconsistent'' voters; and all others. 
\textbf{Right}: Prediction accuracy in each scenario, divided by action type.
}
\end{center}
\end{figure}

Figure~\ref{fig:VotersConsistencyFMeasure} (left) shows the f-measure of all voters, classified by their consistency type. We can see that  the   left tail of the histogram (i.e., almost all voters with low f-measure) are either ``unjustified'' or ``inconsistent.''

This might suggest that perhaps prediction cannot be significantly improved.  
We thus tested how many of the prediction errors themselves were of dominated/inconsistent actions. This can be seen in Figure~\ref{fig:VotersConsistencyFMeasure} (right). The plaid gray bars represent all prediction errors that cannot be explained away as dominated or inconsistent actions.

We conclude that while most of the prediction errors are indeed due voters that behave inconsistently \emph{sometimes}, most of the actual errors are ``plaid'' especially in scenario~F. Thus there is still room for improvement of our decision models. 



\section{Discussion and Conclusions}
It seems that the Attainability-Utility heuristics explains quite well the behavior of most subjects in the data, except those with inherent inconsistencies in their replies. 
To improve the model we can perform more experiments where we use different utilities for the candidates (different utility gaps, negative utility, etc) and have more than 3 candidates. Those experiments can expose  behavior that do not exist in the  current data.
Interestingly, the $\NN$ black-box model sometimes successfully predicts ``unjustified'' actions, and we can try to understand when is this possible.

\subsection{Parameters distribution in the population}
Since the $\alpha$ and $\beta$ parameters correspond to the natural cognitive inclinations, their distribution can in principle reveal important information on the types of strategic voters in the population. 
Unfortunately, it is hard to make clear patterns from the parameter distribution. Indeed, it seems that the distribution of $\beta$ parmeter is bi-modal with peaks at $15$ and $30$ (see appendix). However this is probably an artifact of the way we select parameters for subjects with a large range of optimal parameters (like truthful voters). More experimentation in different conditions is required if we want to better understand the population structure.

One pattern that does stand out is that $\alpha$ values seem to be much higher in the `small $n$' condition.
 This may suggest that not only the relative attainability matters: when $n$ is low and the voter has a substantial chance to be pivotal, the importance of utility increases.

\bibliographystyle{plain}
%

\clearpage
\appendix
\section{Additional Results}

\begin{table}[ht!]
  \begin{center}
    \begin{tabular}{c||c|c|c|c|c|c|c|}
      scenario & $\AVU$ & $\LD$ & $\LD$ + LB & $\CV$ & $\KP$ & $\BL$& $\NN$ \\
      \hline
      A & \textbf{0.902} & \textbf{0.902} & \textbf{0.902} & \textbf{0.902} & \textbf{0.902} &  \textbf{0.902} & \textbf{0.918}   \\
      B & 0.903 & 0.903 & 0.903 & 0.903 & 0.903 &  0.903 & 0.922 \\
      C & 0.870 & 0.389 & 0.819 & 0.389 & 0.703 &  0.693 & 0.741 \\
      D & 0.784 & 0.657 & 0.719 & 0.657 & 0.730 & 0.657 & 0.733 \\
      E & 0.813 & 0.676 & 0.795 & 0.730 & 0.766 & 0.659 & 0.744 \\
      F & 0.728 & 0.525 & 0.609 & 0.533 & 0.640 &  0.411 & 0.638 \\
      \hline
      total & 0.841 & 0.671 & 0.795 & 0.708 & 0.781 & 0.720 & 0.784 \\
    \end{tabular}
    \caption{\label{tab:upper}Upper bounds on the performance of each model. 
    }
  \end{center}
\end{table}

\begin{table}[ht!]
  \begin{center}
    \begin{tabular}{c|c|c|c|c|c}
     & \multicolumn{3}{c}{f-measure}\\
      & total & scenario~C & scenario~D & scenario~E & scenario~F\\
      \hline
      $n < 10$ & 0.775 & 0.747 & 0.767 & 0.735 & 0.534
                          \\\hline
      $n \approx 100$ & 0.739 & 0.710 & 0.653 & 0.746 & 0.588
                          \\\hline
      $n \approx 1000$ & 0.735 & 0.703 & 0.705 & 0.703 & 0.598
                          \\\hline
      $n \approx 10000$ & 0.786 & 0.784 & 0.707 & 0.782 & 0.553\\
    \end{tabular}
    \caption{\label{tab:pollSizeFmeasure}f-measure for different poll size $n$. }
  \end{center}
\end{table}







\begin{figure}[h]
\begin{center}
\includegraphics[scale=1]{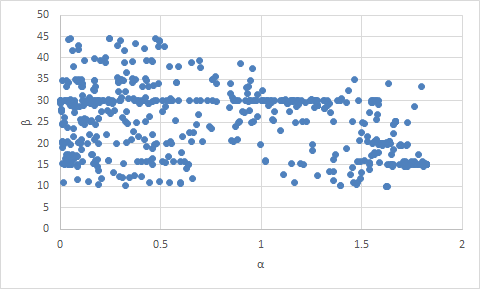}\\
\caption{\label{fig:AlphaVsBeta}A scatter plot ot the $\alpha$ and $\beta$ parameters of each of the 595 voters. }
\end{center}
\end{figure}

\begin{figure}[h]
\begin{center}
\includegraphics[scale=0.8]{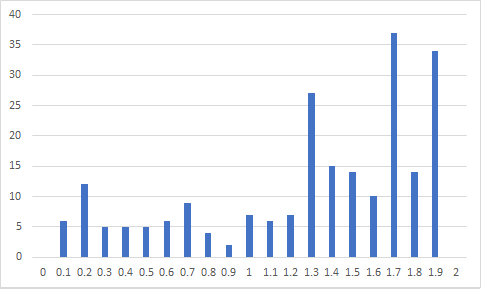}~~
\includegraphics[scale=0.8]{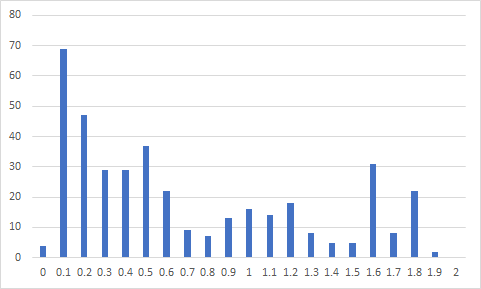}
\caption{\label{fig:alphaDist8}Distribution of the  $\alpha$ parameter for $n < 10$ (left) and $n\geq 100$ (right). }
\end{center}
\end{figure}

\begin{figure}[h]
\begin{center}
\includegraphics[scale=0.75]{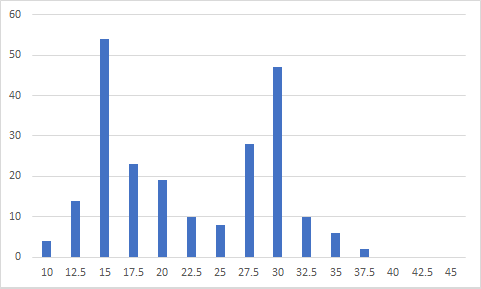}~~
\includegraphics[scale=0.75]{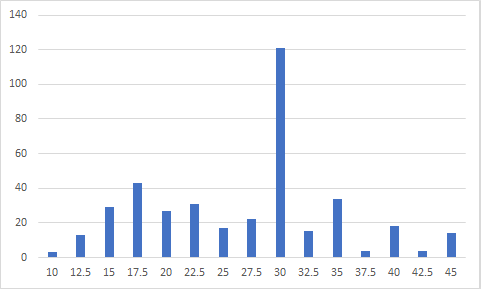}\\
\caption{\label{fig:betaDist8}Distribution of the  $\beta$ parameter for $n <10$ (left) and $n\geq 100$ (right). }
\end{center}
\end{figure}

\section{Black box classifier}
The ``decision model'' $M^\NN$ is a single-hidden-layer feed-forward neural network classifier.
The model consists of a layer of input nodes that represents the features we use in the dataset, A layer of nodes which is called the "hidden" layer
and an output layer that uses the softmax activation function in order to output a classification to one of the possible classes in $C$. The flow of the data is from input to output in one direction and no recurrences occur in the flow process as opposed to recurrent neural networks.
The input nodes are connected to the output nodes via the "hidden" layer nodes (or neurons) by weighted edges. The algorithm is a supervised learning algorithm that takes a set of class labled records and iteratively learns and adjusts the weights on the edges by comparing
the output of each input record to it's class label. The model is iterative and can be updated easily by new records that it haven't seen yet. We use a configuration of 3 units in the "hidden" layer.
In our domain we use a set of vote records that consists of raw and generated features. Some of the generated features are normalized by the number of votes in the poll configuration.
The class label of each record is the selected preference of the voter which can be one of $\{Q,Q',Q''\}$. Using feature selection techniques we selected the following features:
\begin{description}
	\item (a) \textbf{Poll and preference information:}  Candidates poll votes, normalized poll gaps between candidates, preference order, the normalized gap between the leader and the most preferred candidate and the scenario which is the combination of the preference order and the poll information.   
    \item (b) \textbf{Voter information:} A-ratios which are the number of rounds the voter selected action A divided by the number of rounds it was available. A is the action which we determine using the selected preference ($Q,Q'$ or $Q''$) and the order of the preferences in the poll (namely the scenario). We also use the voter type feature which can be one of \{TRT, LB, OTHER\} and is determined by a threshold values over the A-ratio values. For example: if TRT-ratio $>0.9$ then the voter type is TRT (\textit{Truthful}). 
\end{description}
\rmr{Adam can you add details here? explain the features we used and what the algorithm does. } \al{Done here. any comments?}\rmr{It's very good}

	\begin{contact}
		Roy Fairstein\\
		Ben Gurion University\\
		Israel\\
        \email{royfa@post.bgu.ac.il}
	\end{contact}
    
	\begin{contact}
		Adam Lauz\\
		Ben Gurion University\\
		Israel\\
     \email{lauza@post.bgu.ac.il}
	\end{contact}
    
	\begin{contact}
		Kobi Gal\\
		Ben Gurion University\\
		Israel\\
\email{kobig@bgu.ac.il}
	\end{contact}
    
	\begin{contact}
		Reshef Meir\\
		Technion---Israel Institute of Technology\\
		Technion, Israel\\
		\email{Reshef.Meir@ie.technion.ac.il}
	\end{contact}
\end{document}